\documentclass{jfm}

\usepackage{graphicx}
\usepackage{newtxtext}
\usepackage{newtxmath}
\usepackage{natbib}
\usepackage{comment}
\usepackage{hyperref}
\usepackage{caption}
\usepackage{subcaption}

\hypersetup{
    colorlinks = true,
    urlcolor   = blue,
    citecolor  = black,
}

\newcommand{\RomanNumeralCaps}[1]
\linenumbers
\usepackage{lipsum}

\title[Turbulent transport for wall shear stress fluctuations]{Turbulent transport for wall shear stress fluctuations}
\author{Myoungkyu Lee\aff{1}
\corresp{\email{leemk@uh.edu}},
\and Yongyun Hwang\aff{2}}

\affiliation{\aff{1} Department of Mechanical and Aerospace Engineering, The University of Houston, Texas, USA 
\aff{2} Department of Aeronautics, Imperial College London, London, SW7 2AZ, United Kingdom}

\begin{document}
\maketitle

\begin{abstract}
Statistical structure and the underlying energy budget of wall shear stress fluctuations are studied in both Poiseulle and Couette flows with emphasis on its streamwise component. Using a dimensional analysis and direct numerical simulation data, it is shown that the spectra of streamwise wall dissipation for $\lambda \lesssim 1000 \delta_\nu$ are asymptotically invariant with the Reynolds number ($Re$), whereas those for $\lambda \gtrsim \delta$ decay with $Re$ (here, $\lambda$ is a nominal wall-parallel wavelength, and $\delta_\nu$ and $\delta$ are the viscous inner and outer length scales, respectively). The wall dissipation increases with $Re$ due to the increasing contribution of the spectra at $1000 \delta_\nu \lesssim \lambda \lesssim \delta$. The subsequent analysis of the energy budget shows that the near-wall motions associated with these wall dissipation spectra are mainly driven by turbulent transport and are `inactive' in the sense that they contain very little Reynolds shear stress (or turbulence production). As such, turbulent transport spectra near the wall are also found to share the same $Re$-scaling behaviour with wall dissipation, and this is observed in the spectra of both the wall-normal and inter-scale turbulent transports. The turbulent transport underpinning the increase of wall dissipation with $Re$ is characterised by energy fluxes towards the wall, together with inverse energy transfer from small to large length scales along the wall-parallel directions. 
\end{abstract}

\begin{keywords} 
Authors should not enter keywords on the manuscript, as these must be chosen by the author during the online submission process and will then be added during the typesetting process (see \href{https://www.cambridge.org/core/journals/journal-of-fluid-mechanics/information/list-of-keywords}{Keyword PDF} for the full list).  Other classifications will be added at the same time.
\end{keywords}

{\bf MSC Codes }  {\it(Optional)} Please enter your MSC Codes here
 
\section{Introduction}\label{sec:1}
The precise understanding of how fluctuation energy is transported and dissipated through nonlinear interactions between flow structures across a wide range of scales is an outstanding challenge in the studies of turbulence. In wall-bounded turbulence, the large separation between the inner and outer length scales is a distinctive feature at high Reynolds numbers (\(Re\)), and interactions across these scales are crucially involved in the dynamical and statistical processes of the flow \cite[e.g.][]{Cho2018,Lee.20196fe}. In particular, above the near-wall region, there exist a large number of hierarchically organised energy-containing motions, the size of which varies from their distance from the wall to the outer length scale \cite[]{Townsend.1976,Marusic2019}. With increasing \(Re\), the overall impact of these motions becomes more pronounced. It also manifests as the increased peak values in the streamwise velocity variances and the enhanced dissipation of turbulent kinetic energy \cite[e.g.][]{GRAAFF.2000,Marusic2003,Hwang2024,Pirozzoli2024} although there is a different view proposed recently \cite[see e.g.][]{Chen2021,Chen2022,Monkewitz2022}.
Of particular interest in this study is the fluctuation in wall shear stress, a central property of near-wall turbulence relevant to many engineering applications. 

In his early study, \cite{Townsend.1976} speculated that shear stress on the wall contains a slowly fluctuating part, linked with `inactive' part of energy-containing motions in the logarithmic and outer regions. These inactive motions are defined to be with little Reynolds shear stress due to the boundary condition on the wall, which prevents the formation of the wall-normal velocity. In the last decade, there has been growing evidence on the presence of such inactive motions in the vicinity of the wall \cite[e.g.][]{Hwang.2015,Lee.2015wa,Hwang2016c,Deshpande2020}. Strong correlations have also been consistently observed between the large-scale components of velocity fields across the entire wall-normal locations \cite[e.g.][]{Hoyas.2006,Hutchins2007,Mathis2009}, indicating that they may be the consequence of a transfer of energy from the outer flow to the near-wall flow \cite[][]{Lee.20196fe,Yin2024,Deshpande2024}. More recent research has used observed fluctuations in wall shear stress to propose a unified scaling for velocity variance and to extrapolate velocity fields further from the wall \citep{Smits.2021}. 

Despite the recent progress mentioned above, it remains to be understood how these inactive motions near the wall statistically scale and are dynamically formed, especially at high $Re$. Given their origin, the size of inactive motions (say $l$) is supposed to be much larger than that of energy-containing motions in the near-wall region, such as streaks \cite[]{Kline1967} and quasi-streamwise vortices \cite[]{Jeong1997}: i.e. $l \gg \delta_\nu(\equiv \nu/u_\tau$, where $\nu$ and $u_\tau$ are the kinematic viscosity and friction velocity, respectively). Furthermore, they do not carry wall-normal velocity fluctuations, indicating that the related turbulence production must be negligible. 
Therefore, if pressure transport at such large length scales is assumed to be negligible \cite[][see also \S\ref{sec:3.2}]{Lee.20196fe}, the inactive motions would be driven primarily by turbulent transport near the wall and subsequently contribute to dissipation of the turbulence on the wall. We note that the rate of streamwise dissipation at the wall, $\epsilon_{\mathrm{w}}$, directly relates to the streamwise  wall shear stress fluctuation, ${\tau}_\mathrm{w}'$: 
\begin{equation}
\epsilon_\mathrm{w}=-\nu\langle({\tau}_\mathrm{w}'/\mu)^2\rangle,
\end{equation}
where $\mu$ is fluid viscosity and $\langle \cdot \rangle$ is an average in time and in homogeneous directions. 

\begin{figure}
\centering
  \includegraphics[width=\linewidth]{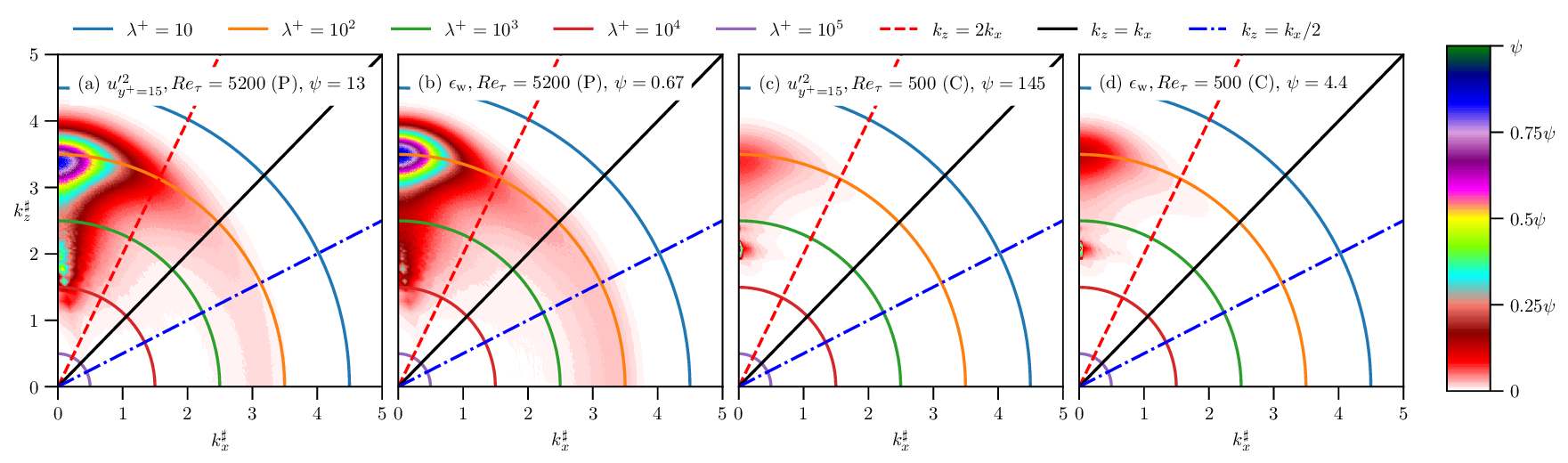}
  \caption{\label{fig:uu15_E_w} Two-dimensional spectral densities in polar-log coordinates of $u'^2$ and $\epsilon_\mathrm{w}$, $k_\mathrm{ref} = 1/ (50000\delta_v)$: 
  (a) $E_{u'^2}^\#$ at $y^+ = 15$, $Re_\tau = 5200$, Poiseuille; 
  (b) $E_{\epsilon_\mathrm{w}}^\#$, $Re_\tau = 5200$, Poiseuille;
  (c) $E_{u'^2}^\#$ at $y^+ = 15$, $Re_\tau = 500$, Couette; 
  (d) $E_{\epsilon_\mathrm{w}}^\#$, $Re_\tau = 500$, Couette. Here, the contour levels are suitably chosen to reveal the similarities between the spectral densities of $u'^2$ and $\epsilon_\mathrm{w}$.}
\end{figure}\label{fig1}

The shear stress fluctuations (or dissipation) on the wall are also related to near-wall velocity fluctuations. In the region close to the wall, the streamwise turbulent kinetic energy (TKE) is approximated as
\begin{equation}\label{eq:1.3}
\langle u'u' \rangle^+ = \epsilon_{\mathrm{w}}^+(y^+)^2+ O((y^+)^3),
\end{equation}
where $u'$ is the streamwise velocity fluctuation  and the superscript `$(\cdot)^+$' denotes normalisation with $u_\tau$ and $\nu$. Figure \ref{fig1} shows the two-dimensional spectral densities of $u'^2$ at $y^+=15$ and $\epsilon_{\mathrm{w}}^+$ in the polar log coordinates (see \S\ref{sec:2} for their definition) for plane Poiseuille flows, driven by a pressure gradient, and Couette flows, driven by boundary movement. Both spectral densities share strong structural similarities, and their importance for Poiseuille flows has been discussed extensively in \cite{Smits.2021}. 
Recently, in relation to (\ref{eq:1.3}), \cite{Chen2021} proposed that the wall dissipation is possibly bounded in the limit of $Re \rightarrow \infty$ and the influence of near-wall bursting on wall dissipation would play a key role in understanding how the peak streamwise turbulence intensity in the near-wall region would scale with $Re$. From this perspective, a precise understanding of the nature of turbulent transport in the near-wall region would illuminate the physical process that underpins the scaling of the near-wall turbulence intensity with $Re$ \cite[][]{Chen2021,Hwang2024,Pirozzoli2024,Jimenez2024} 

The objective of this study is to explore the statistical structure and scaling of turbulent transport associated with wall shear stress fluctuations. The primary focus is given to the streamwise wall shear stress fluctuations. We consider two different forms of wall-bounded turbulence across various $Re$: Poiseuille and Couette flows chosen for their different large-scale behaviours. In Couette flows, the large-scale structures in the outer region are very energetic due to the non-zero turbulence production and mean shear at the channel centre. It was observed that a simulation domain with a streamwise length of $L_x = 100\pi\delta$ ($\delta$ is half height of the channel) did not adequately encompass these structures present in Couette flows at $Re_\tau = 500$ where $Re_\tau$ is the friction Reynolds number \citep{Lee.20181ab}: see also \cite{Hoyas.2024} for this issue at higher $Re$. In contrast, in Poiseuille flows, the production at the channel centre is zero due to vanishing mean shear and Reynolds shear stress at the location. Therefore, the large-scale structures tend to be significantly less energetic than those in Couette flows.

This paper is organised as follows. In \S\ref{sec:2}, we briefly introduce the datasets to be used and the statistical analysis tool (two-dimensional spectra of the energy budget in the polar-log coordinate). The nature of wall dissipation spectra and the related turbulent transport spectra are studied in \S\ref{sec:3} with its scaling behaviour. A summary, a discussion, and implications of the results are finally presented in \S\ref{sec:4}.

\section{Streamwise spectral energy and its budget}\label{sec:2}

The present study uses the results of direct numerical simulations (DNS) of turbulent flows \citep{Lee.2015wa,Lee.20181ab}. These results are produced using the simulation software from \cite{Lee.2013kl}, which employs Fourier-Spectral methods to calculate derivatives in the streamwise ($x$) and spanwise ($z$) directions, along with the seventh-order basis spline methods for derivatives in the wall-normal ($y$) direction: for detailed information on the simulation methodology, the reader may refer to previous studies \citep{Lee.2013kl,Lee.2015wa,Lee.20181ab}. Throughout this study, $u$, $v$, and $w$ indicate the velocity component in the streamwise, wall-normal, and spanwise directions, respectively, and the prime denotes the fluctuations. 


To understand the behaviour in the streamwise wall dissipation and its origin, we consider the following spectral energy balance:
\begin{subequations}\label{eq:3.6}
  \begin{equation}
\frac{\partial E_{u'^2}}{\partial t} =   E_{u'^2}^P+ E_{u'^2}^\Pi+E_{u'^2}^T+ E_{u'^2}^D - E_{u'^2}^{\epsilon}=0,
  \quad \forall (k_x, y, k_z)
  \end{equation}
where  
  \begin{eqnarray}
  E_{u'^2} &=& \langle \hat{u}' \hat{u}'^* \rangle_\mathcal{T}, \\
  E_{u'^2}^P &=& -\langle \hat{u}' \hat{v}'^* +  \hat{u}'^* \hat{v}'\rangle_\mathcal{T} \frac{\partial \langle u \rangle}{\partial y},  \\
  E_{u'^2}^\Pi &=& - \left\langle\hat{u}'\widehat{\frac{\partial p'}{\partial x}}^*+ \hat{u}'^*\widehat{\frac{\partial p'}{\partial x}}\right\rangle_\mathcal{T},  \\
    E_{u'^2}^D &=& \displaystyle \nu \left\langle-k^2 \hat{u}' \hat{u}'^* + \frac{\partial^2 \hat{u}' \hat{u}'^*}{\partial y^2} \right\rangle_\mathcal{T},\\
    E_{u'^2}^T &=& \displaystyle -\left\langle\hat{u}'^* \widehat{\frac{\partial u_k' u'}{\partial x_k}} + \hat{u}'\widehat{\frac{\partial u_k' u'}{\partial x_k}}^* \right\rangle_\mathcal{T}, \\
    E_{u'^2}^\epsilon &=& 2\nu\left\langle\widehat{\frac{\partial u'}{\partial x_k}}\widehat{\frac{\partial u'}{\partial x_k}}^* \right\rangle_\mathcal{T}
  \end{eqnarray}
are the spectral densities of $\langle u'u' \rangle$, its rate of production, streamwise pressure transport, diffusion transport and turbulent transport, and dissipation, respectively. Here, $\langle \cdot \rangle_\mathcal{T}$ denotes the average over time, $(\cdot)^*$ the complex conjugate, and $(\widehat{\cdot})$ indicates the Fourier transform in the $x$ and $z$ directions. 
\end{subequations}

For the two-dimensional spectral density of a statistical quantity as a function of $k_x$ and $k_z$, $E(k_x,k_z)$, the rescaled spectral density in the polar-logarithmic coordinates, $ E^\#$, is used (here, $k_x$ and $k_z$ represent the wavenumbers in the $x$ and $z$ directions, respectively). This form of presentation of the spectral density is introduced to scrutinize the isotropy and the contributions from the $k_x=0$ or $k_z=0$ modes \citep{Lee.20196fe}: 
\begin{subequations}
\begin{equation}\label{eq:2.2a}
  E^\#(k_x^\#, k_z^\#) = k^2 E(k_x,k_z) /\xi,
\end{equation}
where
\begin{equation}
  k_x^\# = \xi k_x / k, \quad  k_z^\# = \xi k_z / k, \quad  k = \sqrt{k_x^2 + k_z^2},
\end{equation}  
\end{subequations}
with $\lambda=2\pi/k$. Here, $k_\mathrm{ref}$ is an arbitrary reference wavenumber, and $\xi = \ln (k/k_\mathrm{ref})$. For inner scaling of the spectra, $k_\mathrm{ref}^+=1/50 000$ and, for outer scaling, $k_\mathrm{ref}\delta=1/100$ are chosen. 
Here, $k_\mathrm{ref}$ determines the distance from the origin in the polar-log coordinates, and $\xi$ is a logarithmic measure of the distance from the origin. Therefore, $k_\mathrm{ref}$ should be smaller than the smallest wavenumber of interest, so that $\xi$ is positive. Also, if $k_\mathrm{ref}$ is too small, it may unnecessarily move the plots away from the origin, making it difficult to see the details of the spectra. In this study, we use $k_\mathrm{ref} = 1/(50000\delta_\nu)$ for inner scaling and $k_\mathrm{ref} = 1/(100\delta)$ for outer scaling, where $\delta_\nu$ and $\delta$ are the viscous inner and outer length scales, respectively.
Note that the integration of $E^\#$ over $k_x^\#$ and $k_z^\#$ remains the same as the integration of $E$ over $k_x$ and $k_z$.

\section{Wall dissipation and turbulent transport}\label{sec:3}

\subsection{Streamwise wall dissipation}\label{sec:3.1}
\begin{figure}
\centering
  \includegraphics[width=\linewidth]{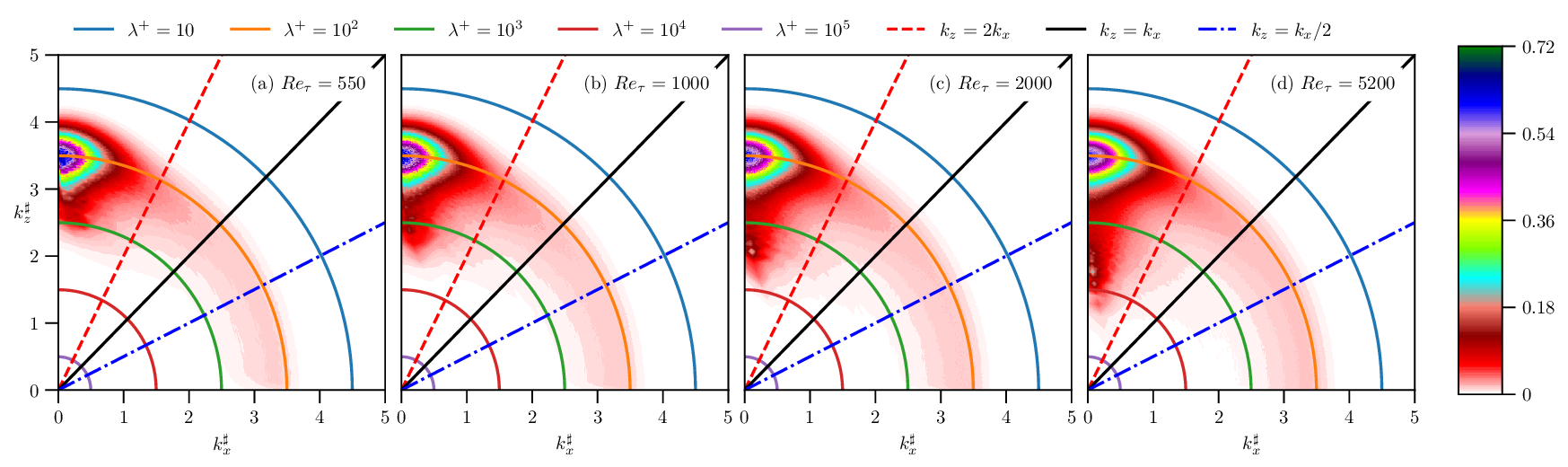}
  \caption{\label{fig:spectra_channel_E_w_inner} Inner-scale normalised two-dimensional spectral densities in polar-log coordinates of $\epsilon_\mathrm{w}$ for Poiseuille flows, $k_\mathrm{ref} = 1/ (50000\delta_v)$.}
\end{figure}

\begin{figure}
\centering
  \includegraphics[width=\linewidth]{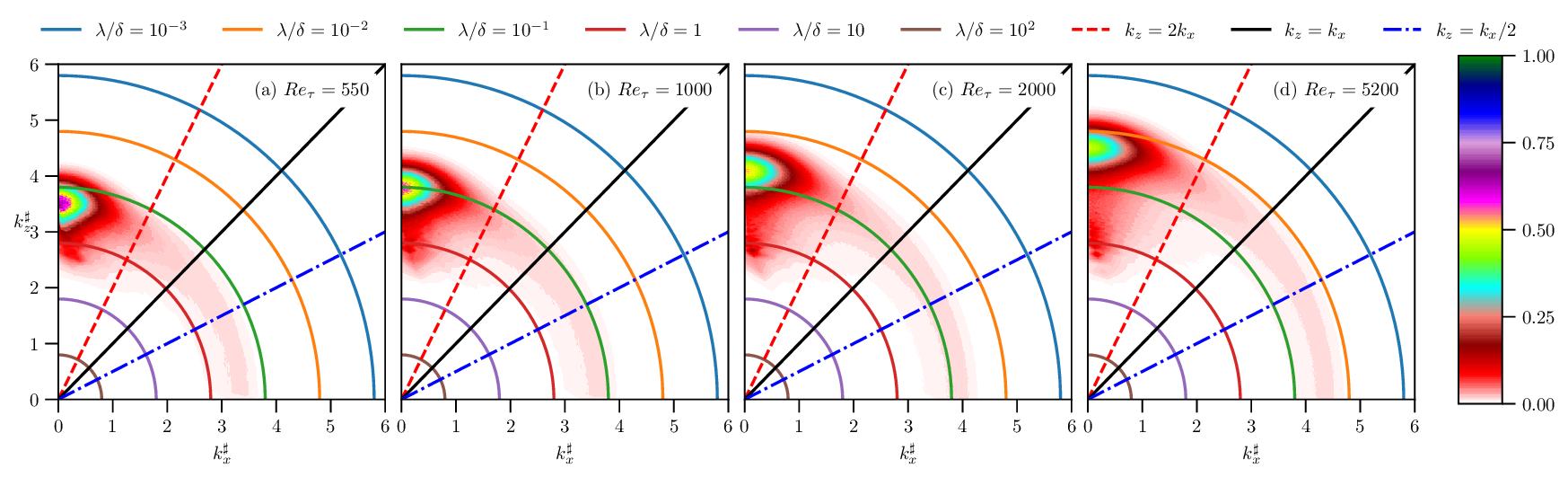}
  \caption{\label{fig:spectra_channel_E_w_outer} Outer-scale normalised two-dimensional spectral densities in polar-log coordinates of $\epsilon_\mathrm{w}$ for Poiseuille flows, $k_\mathrm{ref} = 1/ (100\delta)$.}
\end{figure}

We start by showing the spectral density of the dissipation rate of $\langle u'^2 \rangle$ at the wall, denoted by $E^{\epsilon_\mathrm{w}}_{u'^2}$,
\begin{equation}
  E^{\epsilon_\mathrm{w}}_{u'^2} = E^{\epsilon}_{u'^2}(y = 0)= 2 \nu \left\langle\widehat{\frac{\partial {u}'}{\partial y}}\widehat{\frac{\partial \hat{u}'}{\partial y}}^*\right\rangle_{\mathcal{T},\;y=0},
\end{equation}
in figures~\ref{fig:spectra_channel_E_w_inner} and \ref{fig:spectra_channel_E_w_outer} for Poiseulle flow. 
The pre-multiplied spectral density
of $u'^2$ dissipation, 
$k^2E^{\epsilon_\mathrm{w}}_{u'^2}$ from (\ref{eq:2.2a})
is a function of $k_x$, $k_z$, $u_\tau$, $\nu$ and $\delta$. In the region close to the wall, $\partial/\partial y \sim O(1/\delta_\nu)$ can be assumed. Also, the energy-containing length scale $\delta_\nu$ is identical to the Kolmogorov length scale, becoming the smallest possible length scale. We assume a sufficiently high $Re_\tau$, so that the effect of the mean pressure gradient in Poiseulle flow is negligible in this region -- for example, the effect of the mean pressure gradient on turbulent production becomes less than $1\%$ for $y^+\lesssim 10$ at $Re_\tau \gtrsim 10^3$ (see (\ref{eq:3.5})). For $k_x \sim O(1/\delta_\nu)$ and $k_z \sim O(1/\delta_\nu)$, the Buckingham Pi analysis leads the inner-scaled dimensionless pre-multiplied spectral density of $E^{\epsilon_\mathrm{w}}_{u'^2}$ to be only a function of $k_x^+$ and $k_z^+$: i.e.
\begin{equation}\label{eq:3.2}
\frac{k^2 E^{\epsilon_\mathrm{w}}_{u'^2}(k_x,k_z)}{u_\tau^2 \delta_\nu^{-2}}
= \frac{(k^+)^2 E^{\epsilon_\mathrm{w}}_{u'^2}(k_x^+,k_z^+)}{u_\tau^2 \delta_\nu^{-2}}
=f(k_x^+,k_z^+).
\end{equation}
Figure \ref{fig:spectra_channel_E_w_inner} confirms this analysis, as the spectral density $E^{\epsilon_\mathrm{w}}_{u'^2}$ for $k_x^\#\simeq 0$ and $\lambda^+=100$ (or $k_z^\#=3.5$) is approximately $Re$-invariant. We note that this is also analogous to the behaviour observed for $\langle u'^2 \rangle$ in previous studies \cite[e.g.][see Appendix \ref{appB}]{Hoyas.2006,Lee.2015wa,Lee.20196fe}.

As discussed in \S\ref{sec:1}, the length scale of inactive motions is much larger than that of energy-containing motions in the near-wall region. Given that the largest possible wall-parallel length scale is $\delta$, the pre-multiplied spectral density of $E^{\epsilon_\mathrm{w}}_{u'^2}$ for $k_x \sim O(1/\delta)$ and $k_z \sim O(1/\delta)$ is written as
\begin{equation}\label{eq:3.3}
\frac{k^2 E^{\epsilon_\mathrm{w}}_{u'^2}(k_x,k_z)}{u_\tau^2 \delta_\nu^{-2}}= \frac{(k \delta)^2 E^{\epsilon_{\mathrm{w}}}_{u'^2}(k_x\delta,k_z\delta,\delta/\delta_\nu)}{u_\tau^2 \delta_\nu^{-2}}=g(k_x\delta,k_z\delta;Re_\tau),
\end{equation}
where $Re$-dependence appears due to the presence of the two different length scales, $\delta_\nu$ and $\delta$. A weak $Re$-dependence is confirmed from the outer-scaled spectral intensity in figure \ref{fig:spectra_channel_E_w_outer}: although the spectral density around the peak location at $k_x^\#\simeq 0$ and $\lambda/\delta=1$ (or $k_z^\#\simeq 2.5$) does not exhibit a visible $Re$-dependence, the region around $k_z\simeq 2k_x$ and $\lambda/\delta=1$ ($k_x^\#\simeq 1.5$ and $k_z^\# \simeq 2.5$) shows a gradual weakening in the spectral density with increasing $Re_\tau$. 

\begin{figure}
\centering
  \includegraphics[width=\linewidth]{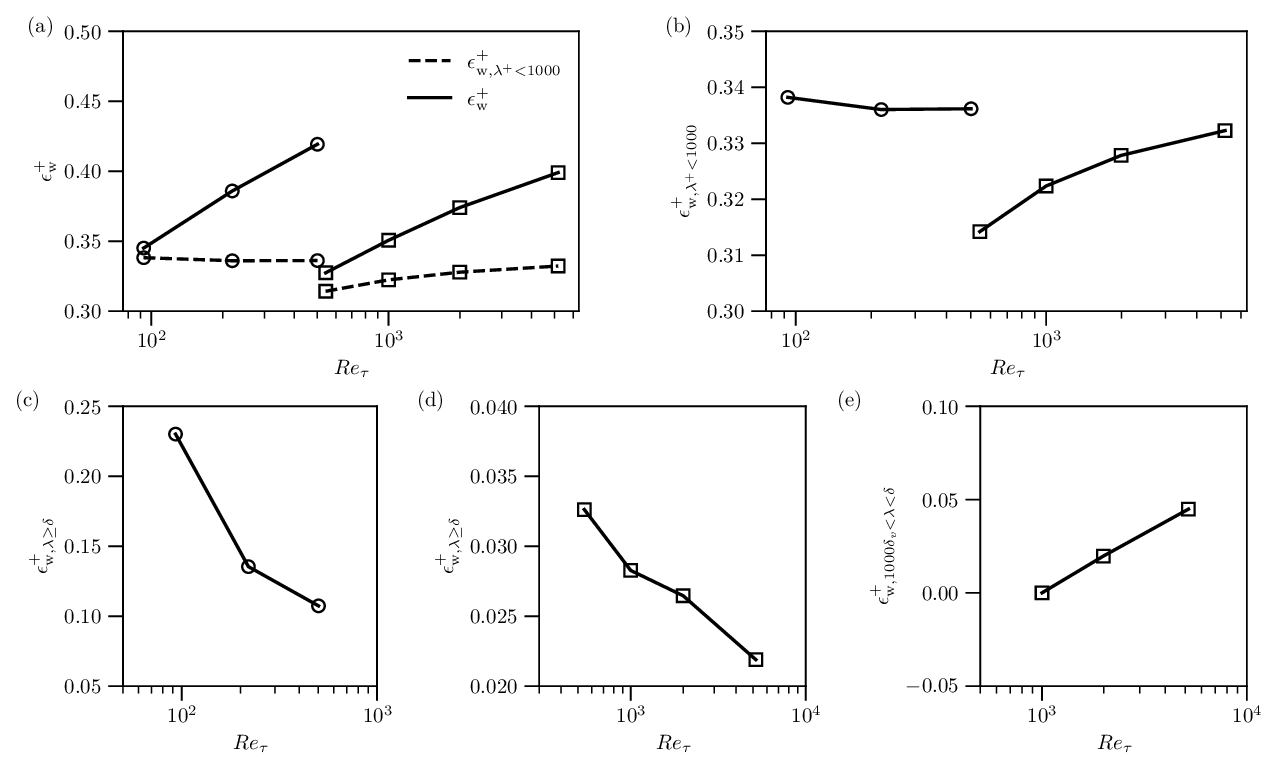}
  \caption{\label{fig:Euu_wall} Dissipation rate of $\langle u'^2 \rangle^+$ at walls, $\epsilon_\mathrm{w}$ for Couettte ($\circ$) and Poiseuille ($\square$) flows at various $Re_\tau$: 
  (a) total and inner-scale-filtered $\epsilon_\mathrm{w}$;
  (b) inner-scale-filtered $\epsilon_\mathrm{w}$, magnification of (a);
  (c) outer-scale-filtered $\epsilon_\mathrm{w}$ for Couette flow;
  (d) outer-scale-filtered $\epsilon_\mathrm{w}$ for Poiseuille flow.
  (e) difference between $\epsilon_\mathrm{w}$ and the sum of inner-scale-filtered $\epsilon_\mathrm{w}$ and outer-scale-filtered $\epsilon_\mathrm{w}$ (Poiseulle flow for $Re_\tau>1000$).
  }
\end{figure}

Figure \ref{fig:Euu_wall} shows the rate of streamwise dissipation on the wall, together with its high- and low-pass filtered values using inner- and outer-scaled cut-off wavelengths, $\lambda^+=1000$ and $\lambda/\delta=1$, respectively. We note that for $Re_\tau <1000$ there is an overlap part between the two filtered quantities. From this viewpoint, these properties are more useful as $Re_\tau$ is sufficiently large. The rate of streamwise dissipation on the wall increases with $Re$, similar to the well-known streamwise turbulence intensity near the wall (figure \ref{fig:Euu_wall}a). When the high-pass filter with $\lambda^+=1000$ is applied (figure \ref{fig:Euu_wall}b), the dissipation rate remains approximately constant in Couette flow at all the Reynolds numbers, consistent with (\ref{eq:3.2}). In Poiseulle flow, it is seen to asymptotically approach the constant value observed in Couette flow. This difference can be explained with the streamwise mean momentum equation of Poiseulle flow given by
\begin{equation}\label{eq:3.4}
\frac{dU^+}{dy^+}-\langle u'v'\rangle^+=\left(1-\frac{y^+}{Re_\tau}\right),
\end{equation}
where the $Re$-dependent last term stems from its mean streamwise pressure gradient. From (\ref{eq:3.4}), the peak value of the rate of turbulence production is obtained as
\begin{equation}\label{eq:3.5}
P^+(y_p^+)= \frac{1}{2}\left(1-\frac{y_p^+}{Re_\tau}\right),
\end{equation}
where $P^+\equiv -2\langle u'v' \rangle^+ dU^+/dy^+$ and $y_p^{+}$ is its peak wall-normal location near the wall. In Couette flow, $P^+(y_p^+)= 1/2$ for all Reynolds numbers due to the absence of the $Re$-dependent term (i.e. the mean streamwise pressure gradient term) in (\ref{eq:3.4}). In the near-wall region, turbulence production occurs for $\lambda^+<1000$ \cite[][ see also figures \ref{fig:filtered_budget_poiseuille} and \ref{fig:filtered_budget_couette}]{Lee.20196fe}. Then, the related dissipation at $\lambda^+<1000$ would be directly affected by the production, implying that it would asymptotically reach a constant value as $Re\rightarrow \infty$, consistent with figure \ref{fig:Euu_wall}(b). 

When the streamwise dissipation rate is low-pass filtered with $\lambda/\delta=1$ (figures \ref{fig:Euu_wall}c,d), the remaining part becomes substantially small. However, the small filtered streamwise dissipation rate decays slowly with $Re$, consistent with the outer-scaled spectral intensity of $E^{\epsilon_\mathrm{w}}_{u'^2}$ in figure \ref{fig:spectra_channel_E_w_outer} and the dimensional analysis in (\ref{eq:3.3}). Given that the dissipation rate for $\lambda^+<1000$ is asymptotically constant (figures \ref{fig:Euu_wall}b) and that for $\lambda/\delta>1$ decays with $Re$ (figures \ref{fig:Euu_wall}c, d), the increase in the rate of total streamwise dissipation on the wall with $Re$ must be driven primarily by $E^{\epsilon_\mathrm{w}}_{u'^2}$ for $1000 \delta_\nu<\lambda<\delta$ at high $Re$ (figure~\ref{fig:Euu_wall}e). 
Note that the range $1000 \delta_\nu < \lambda < \delta$ exists only for sufficiently high $Re$ (specifically, $Re_\tau > 1000$). For such cases, the contribution to wall dissipation from this intermediate range can be expressed as $\epsilon_{\mathrm{w},\,1000 \delta_\nu < \lambda < \delta}^+ = \epsilon_{\mathrm{w}}^+ - \epsilon_{\mathrm{w},\,\lambda^+ < 1000}^+ - \epsilon_{\mathrm{w},\,\lambda \ge \delta}^+$, and the increase of $\epsilon_{\mathrm{w}}^+$ from $Re_\tau=1000$ to $Re=5000$ in Poiseulle flow is almost identical to that of $\epsilon_{\mathrm{w},\,1000 \delta_\nu < \lambda < \delta}^+$.
By definition, this is the length scale associated with the logarithmic layer, supporting the idea of \cite{Townsend.1976} that shear stress on the wall contains a fluctuating part associated with energy-containing motions in the log and outer regions. We also note that the same behaviour was also observed from the near-wall streamwise velocity variances \cite[][see also Appendix \ref{appB}]{Hwang2024,Pirozzoli2024}, consistent with the relation (\ref{eq:1.3}).

\subsection{Streamwise spectral energy budget in near-wall region}\label{sec:3.2}

\begin{figure}
\centering
  \includegraphics[width=0.9\linewidth]{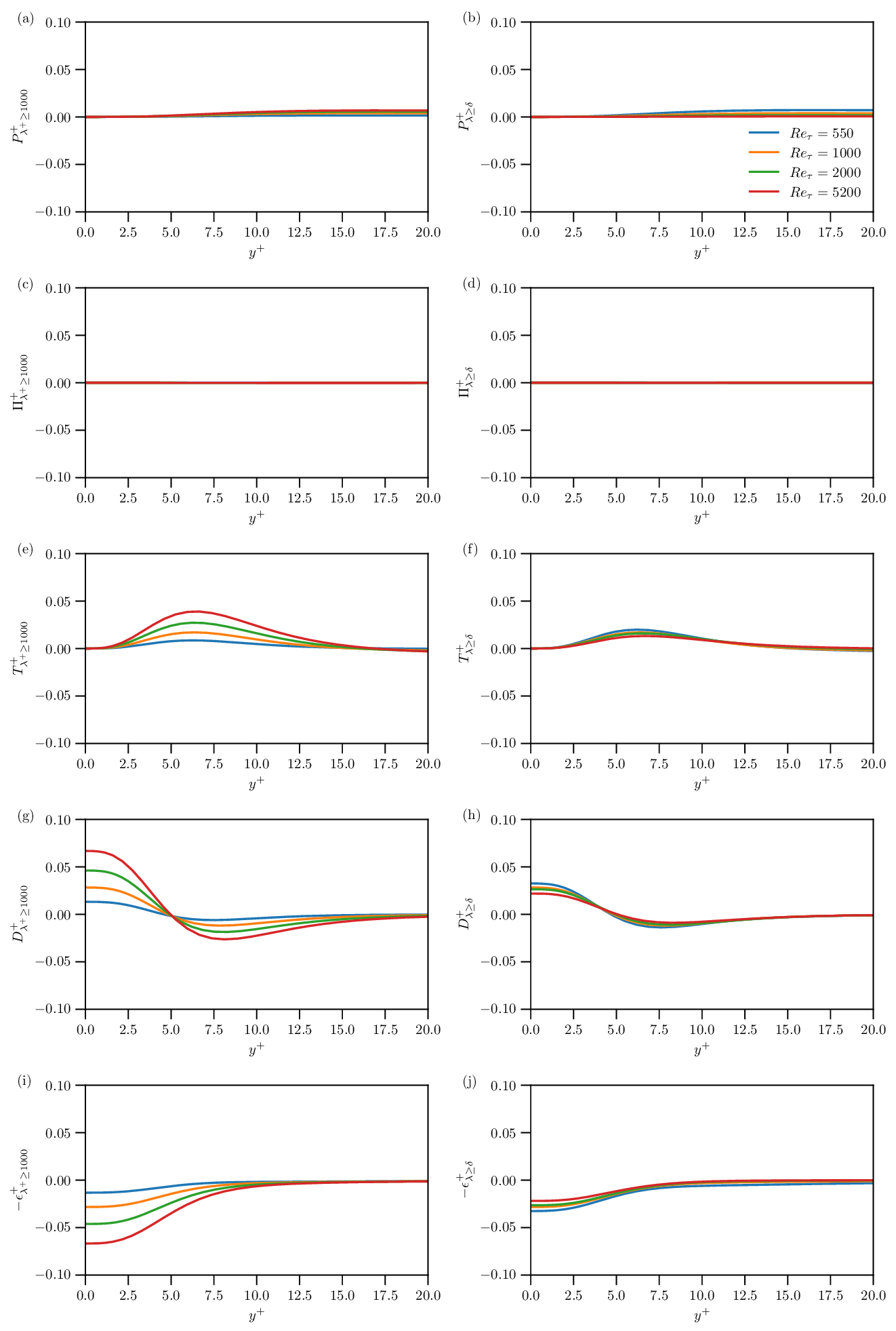}
  \caption{\label{fig:filtered_budget_poiseuille} Filtered budget terms near the wall of Poiseuille flows. (a,b) Production; (c,d) pressure transport; (e,f) turbulent transport; (g,h) viscous transport; (i,j) dissipation rate.}
  \end{figure}

\begin{figure}
\centering
  \includegraphics[width=0.9\linewidth]{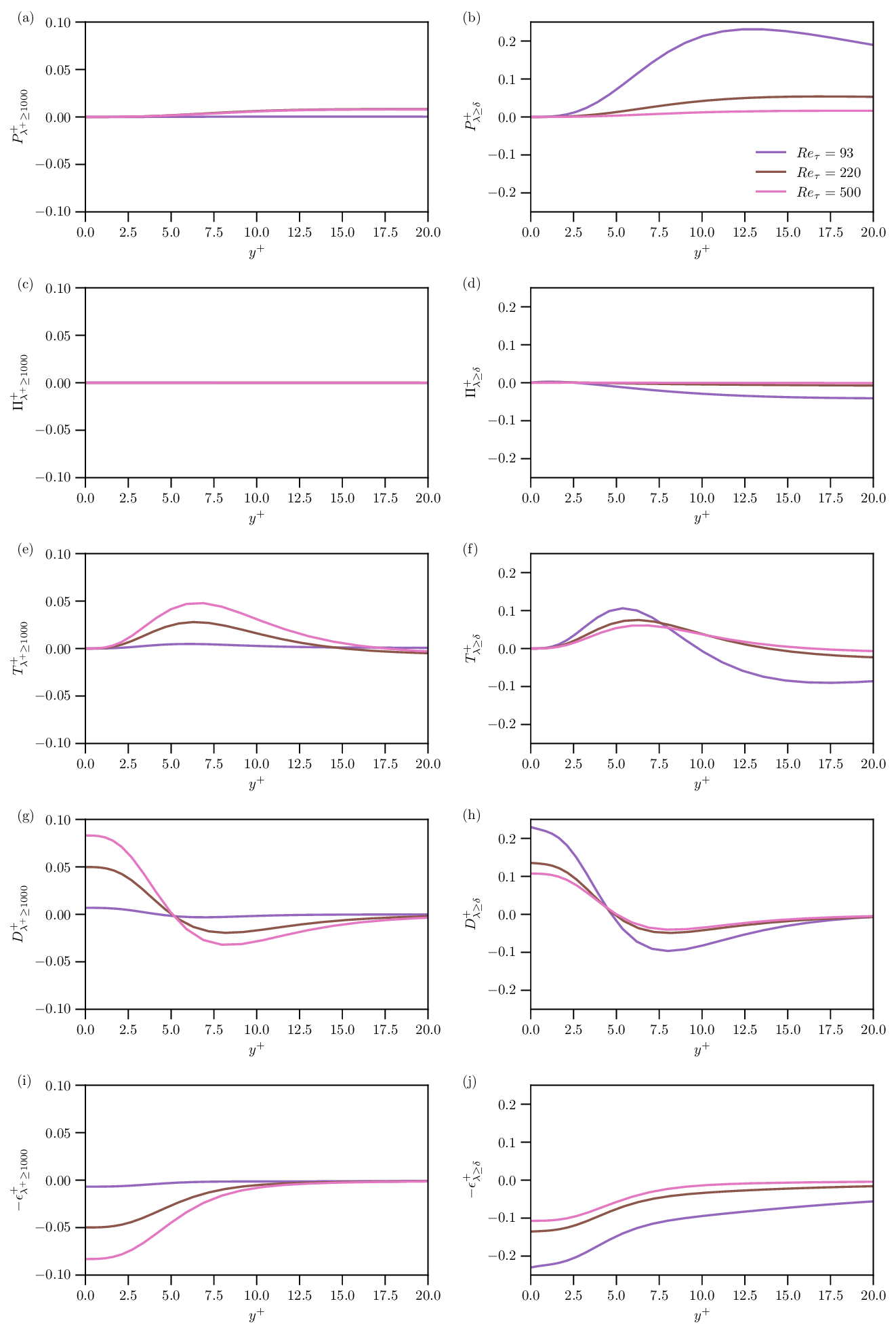}
  \caption{\label{fig:filtered_budget_couette} Filtered budget terms near the wall of Couette flows. (a,b) Production; (c,d) pressure transport; (e,f) turbulent transport; (g,h) viscous transport; (i,j) dissipation rate.}
  \end{figure}

Now, we consider two low-pass filters with the cut-off wavelengths given by $\lambda^+=1000$ and $\lambda/\delta=1$. Their applications to each of the energy budget terms in (\ref{eq:3.6}) are shown in figures \ref{fig:filtered_budget_poiseuille} and \ref{fig:filtered_budget_couette} for the near-wall region. When $\lambda^+=1000$ is considered for the cut-off wavelength (left column of figure \ref{fig:filtered_budget_poiseuille} and \ref{fig:filtered_budget_couette}), the rates of turbulence production ($P^+_{\lambda^+\geq 1000}$) and pressure transport ($\Pi^+_{\lambda^+\geq 1000}$) are negligibly small, compared to the other budget terms. Therefore, in the near-wall region (say $y^+ \lesssim 10$), the following energy balance is obtained for the spectral density of streamwise TKE at $\lambda^+\geq 1000$:
\begin{equation}\label{eq:3.7}
T^+_{\lambda^+\geq 1000}+D^+_{\lambda^+\geq 1000}-\epsilon^+_{\lambda^+\geq 1000}\simeq 0,
\end{equation}
where $T$, $D$ and $\epsilon$ are the rates of streamwise turbulent transport, viscous diffusion transport and dissipation, respectively. This energy balance is consistent with the concept of `inactive' motions by \cite{Townsend.1976}, as these large streamwise motions near the wall do not contain Reynolds shear stress (or turbulence production, equivalently). More importantly, they are driven by turbulent transport, $T$. Furthermore, given that turbulence transport must vanish at the wall (i.e. $T^+_{\lambda^+\geq 1000}=0$ at $y=0$), (\ref{eq:3.7}) becomes
\begin{equation}
D^+_{\lambda^+\geq 1000}-\epsilon^+_{\lambda^+\geq 1000}=0
\end{equation}
at $y=0$. Since $T_{\lambda>1000}^+>0$ in the near-wall region, this acts as the source term for non-zero $D_{\lambda>1000}^+$ and $\epsilon_{\lambda>1000}^+$ (see figures \ref{fig:filtered_budget_poiseuille}c  and \ref{fig:filtered_budget_couette}c). 
We note $\int_\Omega D_{\lambda>1000}^+ dy=0$, where $\Omega$ is the wall-normal domain. Therefore, this suggests that the energy of inactive motions driven by turbulent transport near the wall is transported to the wall by viscous diffusion, which ultimately contributes to dissipation on the wall for $\lambda^+\geq 1000$.
The same energy balance is obtained by low-pass filtering with $\lambda/\delta=1$ as long as $Re_\tau$ is sufficiently high (right column of figures \ref{fig:filtered_budget_poiseuille}  and \ref{fig:filtered_budget_couette}). However, for $Re_\tau < 1000$, the outer-scaled filtered wavelength $\lambda=\delta$ becomes $\lambda^+<1000$. Therefore, the filtered rate of production ($P^+_{\lambda\geq \delta}$) and the pressure transport ($\Pi^+_{\lambda\geq \delta}$) are no more negligible due to the contribution from the energy-containing motions in the near-wall region. 

In figures \ref{fig:filtered_budget_poiseuille}  and \ref{fig:filtered_budget_couette}, all the terms in (\ref{eq:3.7}) increase with $Re$, consistent with the observation in figure \ref{fig:Euu_wall}, where the increase of the streamwise wall dissipation with $Re$ primarily originates from $\epsilon^+_{\lambda^+\geq 1000}$. Since the driving term in (\ref{eq:3.7}) is turbulent transport, we explore its two-dimensional spectral density near the wall further by decomposing into
\begin{subequations}\label{eq:4}
  \begin{eqnarray}
    E_{u'^2}^{T^\bot} &=& -\frac12 \left\langle  \frac{\partial \hat{u}'^* \widehat{u'v'}}{\partial y} + \frac{\partial \hat{u}' \widehat{u'v'}^*}{\partial y} \right\rangle_\mathcal{T}, \\
    E_{u'^2}^{T^\|} &=& E_{u'^2}^{T} - E_{u'^2}^{T^\bot}.
  \end{eqnarray}    
\end{subequations}
Here, $\int_0^{2h} E_{u'^2}^{T^\bot} \mathrm{d}y = 0$ for all $(k_x,k_z)$, such that there is no net turbulent transport in the wall-normal direction by $E_{u'^2}^{T^\bot}$. Similarly, $\iint_0^{\infty} E_{u'^2}^{T^\|} \mathrm{d}k_x \mathrm{d}k_z = 0$ for all $y$, and there is no net transfer between scales by $E_{u'^2}^{T^\|}$ at any given wall-normal distance. Hence, $E_{u'^2}^{T^\bot}$ represents turbulent transport in the wall-normal direction \cite[e.g. energy transport from outer flow;][]{Lee.20196fe}, while $E_{u'^2}^{T^\|}$ signifies the inter-scale energy transfer at given wall-normal distances \cite[e.g. inverse energy transfer from small to large scale;][]{Cho2018,Kawata.2018,Lee.20196fe}.

\begin{figure}
\centering
  \includegraphics[width=\linewidth]{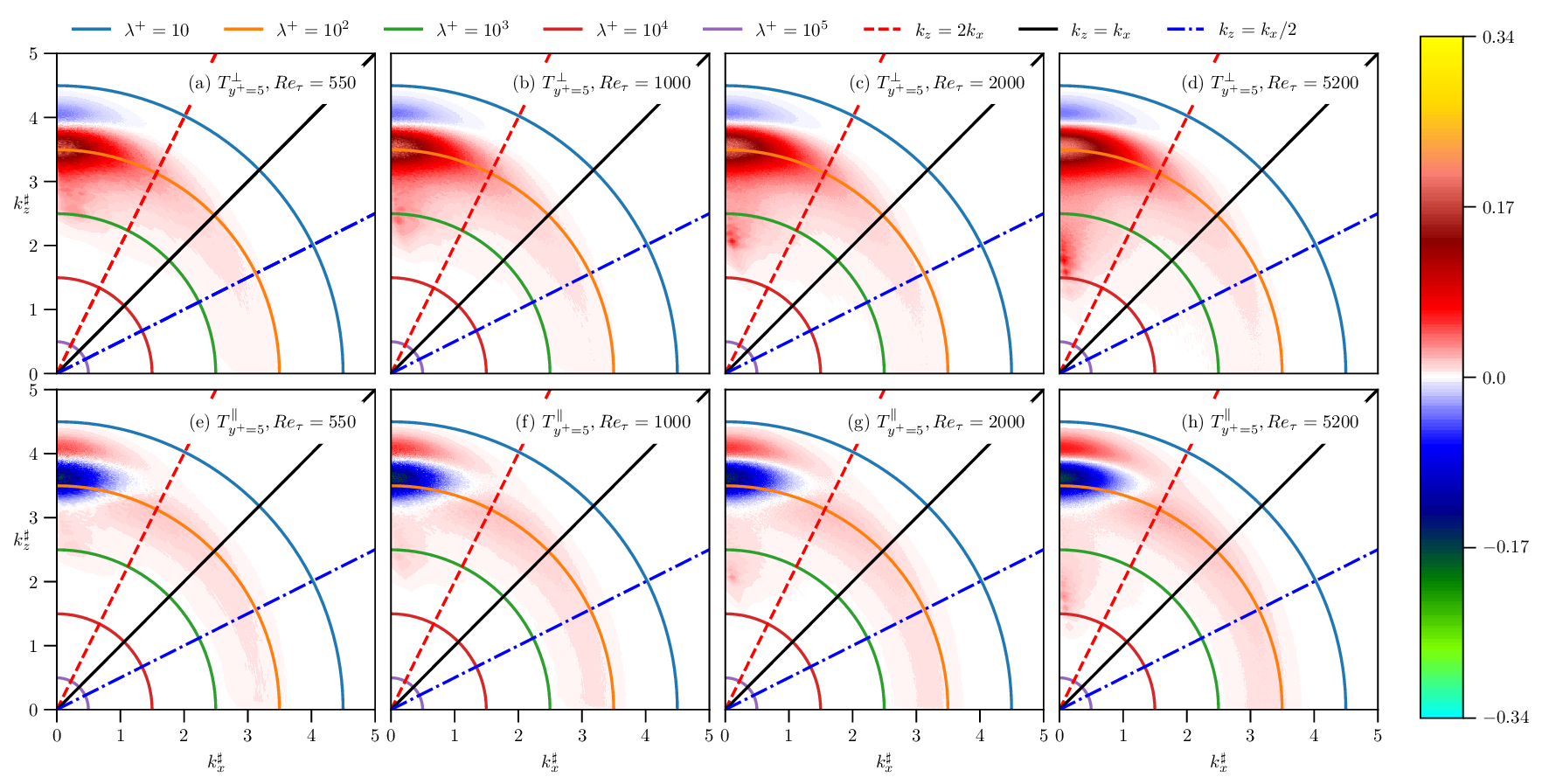}
  \caption{\label{fig:T_spectra_channel_inner} 
  Inner-scale normalised two-dimensional spectral densities of wall-normal transport and inter-scale transfer  of $\langle u'^2\rangle$ in polar-log coordinates, $y^+ = 5$, Poiseuille flows, $k_\mathrm{ref} = 1/ (50000\delta_v)$: (a-d) wall-normal transport, $T^\bot$; (e-h) inter-scale transfer $T^\|$.}
\end{figure}

\begin{figure}
\centering
  \includegraphics[width=\linewidth]{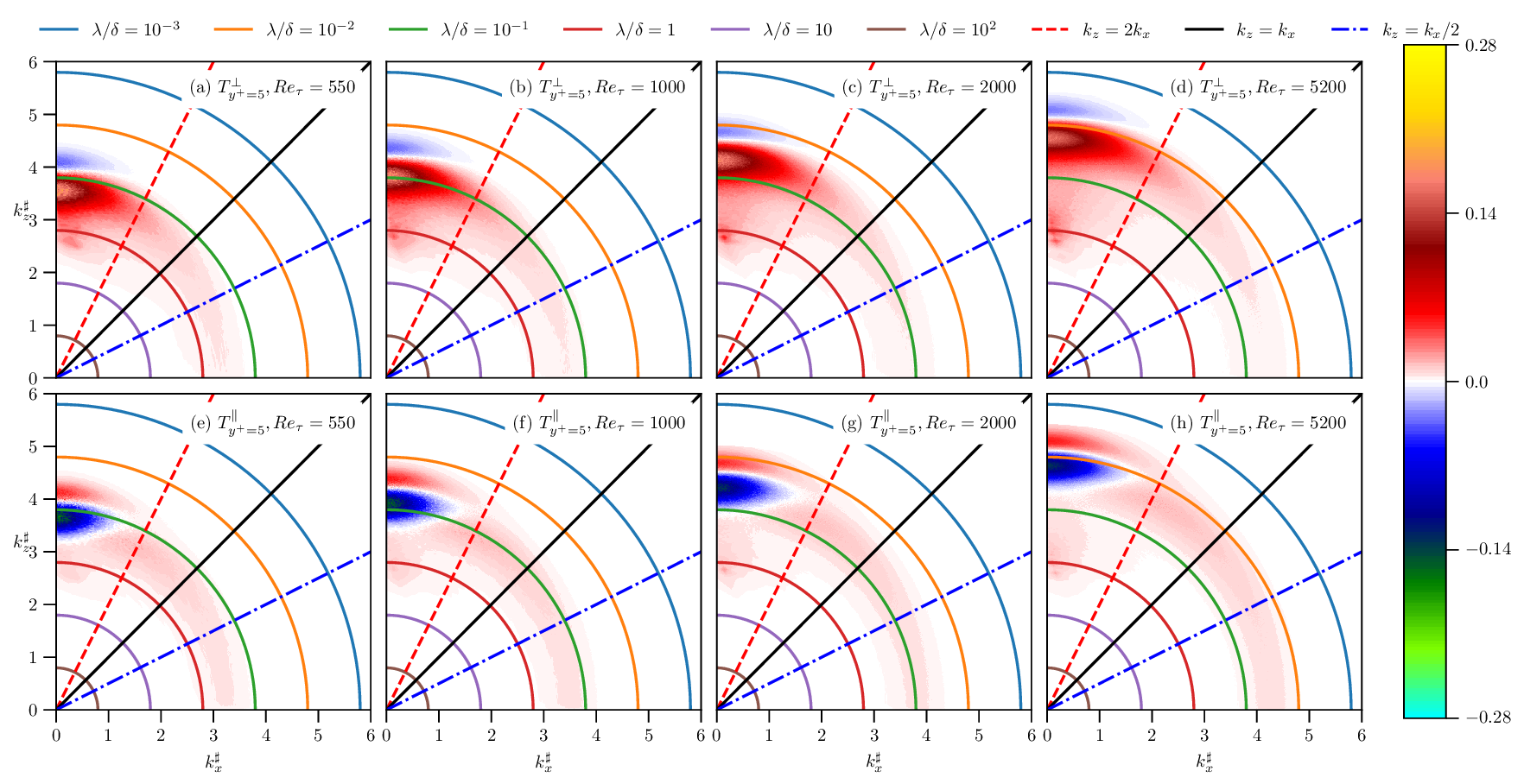}
  \caption{\label{fig:T_spectra_channel_outer}
    Outer-scale normalised two-dimensional spectral densities of wall-normal transport and inter-scale transfer of $\langle u'^2\rangle$ in polar-log coordinates, $y^+ = 5$, Poiseuille flows, $k_\mathrm{ref} = 1/ (100\delta)$: (a-d) wall-normal transport, $T^\bot$; (e-h) inter-scale transfer $T^\|$.}
\end{figure}

\begin{figure}
\centering
  \includegraphics[width=\linewidth]{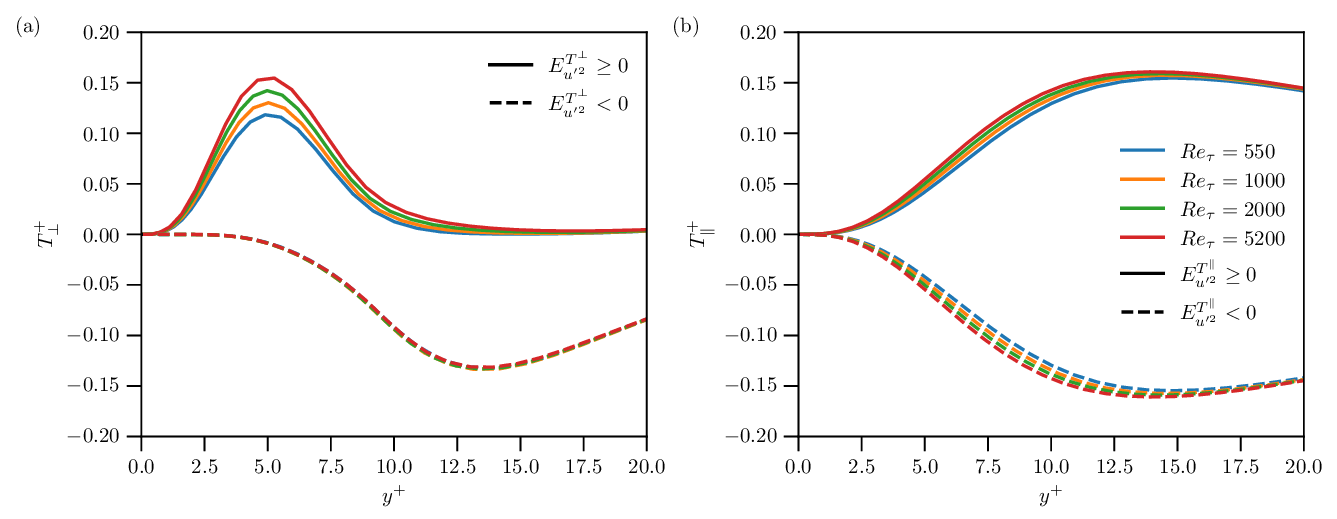}
  \caption{\label{fig:T_conditional}
  Turbulent transport near the wall of Poiseulle flow, conditioned with the positive and negative part of the corresponding spectra: (a) wall-normal transport $T^\bot$; (b) inter-scale transfer $T^\|$.}
\end{figure}

Figures~\ref{fig:T_spectra_channel_inner} and \ref{fig:T_spectra_channel_outer} show the inner- and outer-scaled spectral densities of $E_{u'^2}^{T^\bot}$ and $E_{u'^2}^{T^\|}$ at $y^+ = 5$ in Poiseulle flow. The positive (red) peak region of $E_{u'^2}^{T^\bot}$ appears at $\lambda^+\approx 100$ and $k_x^\#\approx 0$ (figures~\ref{fig:T_spectra_channel_inner}a-d), indicating the wall-normal energy transport from the elongated streaks (i.e. positive/negative streamwise velocity fluctuations elongated in the streamwise direction) in the buffer layer. The adjacent negative (blue) peak region at $\lambda^+<100$ must imply an outward energy transport, given that the spectral density of the streanwise wall dissipation in this region is very small (see figure \ref{fig:spectra_channel_E_w_inner}). In the spectra of $E_{u'^2}^{T^\|}$ (figures~\ref{fig:T_spectra_channel_inner}e-h), a negative (blue) peak region appears at $\lambda^+\approx 100$ and $k_x^\#\approx 0$, characterising the wall-parallel energy transfer from the energy-containing length scale to the other length scales in the near-wall region. The positive (red) peak region in the same spectra appears at $\lambda^+< 100$ and $k_x^\#\approx 0$. We note that this peak value is comparable to the magnitude of the negative peak in the $E_{u'^2}^{T^\bot}$ spectra. Given that these two peaks occur on a length scale smaller than $\lambda^+=100$ (i.e. classic energy cascade) and involve an outward energy transport, they would statistically represent the near-wall bursting \cite[]{Kline1967} for the streamwise TKE. 

The inner-scaled spectral densities of $E_{u'^2}^{T^\bot}$ associated with the outward energy transport appear to be approximately $Re$-independent (figures~\ref{fig:T_spectra_channel_inner}a-d) -- given the analysis in \S\ref{sec:3.1}, even if there exist some variations with $Re$, it is expected that the extent will be as small as $\epsilon^+_{\lambda^+<1000}$ relative to $\epsilon^+$ in figure \ref{fig:Euu_wall}. This is confirmed in figure \ref{fig:T_conditional}, where the the wall-normal and inter-scale turbulent transport are obtained by conditionally sampling positive and negative parts of the corresponding spectra shown in figure~\ref{fig:T_spectra_channel_inner}. Indeed, the approximate $Re$-independent behaviour of turbulent transport is only observed from $T_\bot^+$ associated with the loss of turbulent kinetic energy (either to the near-wall and outer region; dashed lines in figure \ref{fig:T_conditional}a). \cite{Chen2021} recently proposed that the bursting process in the near-wall region would cause the wall dissipation to vary with $Re$, and this idea was central in the development of their asymptotic theory for scaling the peak streamwise variances with $Re$ in the near-wall region. However, the bursting-related turbulent transport ($\lambda^+<100$ in figure~\ref{fig:T_spectra_channel_inner} and dashed lines in figure \ref{fig:T_conditional}a) do not support this idea, as they are approximately independent of $Re_\tau$. In fact, the spectra of $E_{u'^2}^{T^\bot}$ and $E_{u'^2}^{T^\|}$ depend on $Re$ most strongly for $\lambda^+> 1000$ like the wall dissipation spectra in figure \ref{fig:spectra_channel_E_w_inner}. This involve the wall-normal energy gain through the wall-normal transport and the inter-scale energy transport, as shown in figure \ref{fig:T_conditional} (see below for a further discussion). Furthermore, when the spectra of $E_{u'^2}^{T^\bot}$ and $E_{u'^2}^{T^\|}$ are scaled by outer units, their $Re$-dependent behaviour is qualitatively similar to that of wall dissipation in figure \ref{fig:spectra_channel_E_w_outer} -- a small positive peak appears at $\lambda/\delta\approx 1$ and $k_x^\#\approx 0$, and the spectral densities for $\lambda/\delta\approx 1$ and $k_z \approx 2k_s$ slowly decay with $Re$ (figure~\ref{fig:T_spectra_channel_outer}).

\begin{figure}
\centering
  \includegraphics[width=0.95\linewidth]{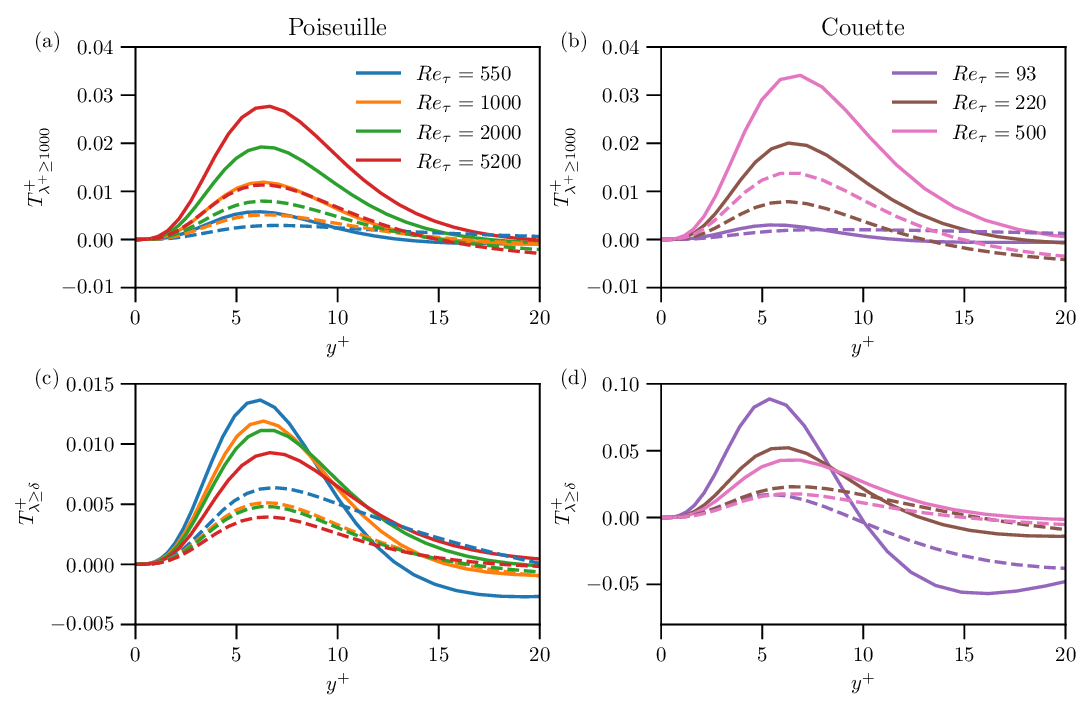}
  \caption{\label{fig:T_filtered} Turbulent wall-normal transport, $T^+_\bot$, and inter-scale transfer, $T^+_\|$ of $\langle u'^2 \rangle^+$:  (------) $T^+_\bot$,  (-- -- --) $T^+_\|$. (a-b) Inner-scale filtered with $\lambda^+=1000$; (c-d) outer-scale filtered with $\lambda=\delta$.}
\end{figure}

To further understand the nature of near-wall turbulent transport, $T_\bot$ and $T_\|$ obtained by integrating the related spectra for $\lambda^+\geq 1000$ and $\lambda \geq \delta$ are also shown in figure \ref{fig:T_filtered}. For both Poiseulle and Couette flows, the values of $T_\bot$ and $T_\|$ increase with $Re_\tau$, when the spectra of $\lambda^+\geq 1000$ are considered. In contrast, when $\lambda \geq \delta$ is considered, the values of $T_\bot$ and $T_\|$ decrease with $Re_\tau$. In all the cases considered here, the near-wall turbulent transport is dominated by the wall-normal one, $T_\bot$, as it is found to be two to three times greater than its inter-scale counterpart, $T_\|$. Given that $T_\bot$ is predominantly positive below $y^+\simeq 10$ (figure \ref{fig:T_filtered}), this transport originates mainly from $y^+\geq 10$.

\section{Discussions}\label{sec:4}

Thus far, we have explored the scaling behaviour of wall shear stress fluctuations in both Poiseulle and Couette flows through the rate of wall dissipation, focusing on its streamwise component. A dimensional analysis suggests the inner-scaling nature of the spectra of streamwise wall dissipation for $\lambda \sim O(\delta_\nu)$ and a $Re$-dependent behaviour for $\lambda \sim O(\delta)$. This was confirmed by DNS data; in particular, the rate of streamwise wall dissipation only accounting for $\lambda^+ \leq 1000$ is approximately constant (Couette flow) or reaches the constant asymptotically on increasing $Re$ (Poiseulle flow), while the one for $\lambda > \delta$ consistently decays with $Re$. Subsequently, an energy budget analysis was performed to relate this observation to the large streamwise motions near the wall ($y^+\leq 10$), which are inactive in the sense that they do not involve Reynold shear stress or turbulence production. It was found that these motions are directly driven by near-wall turbulent transport, which exhibits the same $Re$-scaling behaviour with the wall dissipation. This was also observed for both the wall-normal and inter-scale turbulent transport at all the Reynolds numbers considered in Couette and Poiseulle flows.

Given that the $Re$-scaling behaviours of the streamwise wall dissipation spectra originate from turbulent transport spectra near the wall, it is useful to investigate the turbulent transport a little further. In the Fourier space, the spectral density of turbulent transport is written as a convolution integral across the entire wavenumbers, i.e.
\begin{equation}\label{eq:4.1}
E_{u'^2}^{T}(k_x,k_z) \sim 
\hat{u}^*(k_x,k_z)\iint_{-\infty}^{\infty}\hat{u}_j(k_x-s_1,k_z-s_2)\frac{\partial \hat{u}(s_1,s_2)}{\partial x_j} \textrm{d}s_1\textrm{d}s_2,  
\end{equation}
and similar expressions may be written for both $E_{u'^2}^{T^\bot}$ and $E_{u'^2}^{T^\|}$, given their definitions in (\ref{eq:4}). 

A triadic wave interaction form of turbulent transport in (\ref{eq:4.1}) suggests some possible origins for the $Re$-dependent behaviours of the large streamwise motions inactive in the near-wall region. First, given their large size relative to the wall-normal location of interest ($y^+ \lesssim 10$), these motions are expected to develop a `boundary layer' near the wall. In other words, $\hat{u}_j$ for $\lambda^+ > 1000$ near the wall would be a function of $Re$ to satisfy the no-slip boundary condition as in Prandtl's laminar boundary layer \cite[e.g.][]{Schlichting2017}. The degree to which each of $\hat{u}_j$ feels the wall would depend on its size (or horizontal wavelengths), since their length scales vary in the range of $O(\delta_\nu) \ll l \lesssim O(\delta)$. For the same reason, the time scale of $\hat{u}_j$ for each $k_x$ and $k_z$ would also matter in characterising the thickness of the boundary layer as in the Stokes second problem \cite[e.g.][]{Schlichting2017}. However, perhaps more importantly, the form of (\ref{eq:4.1}) implies that each of $\hat{u}_j$ is also a consequence of triadic interactions across all wavenumbers. Given that the length scale near the wall varies from $O(\delta_\nu)$ to $O(\delta)$, the most evident $Re$-dependence of (\ref{eq:4.1}) would be given in its integration limit, such that 
\begin{equation}\label{eq:4.2}
E_{u'^2}^{T}(k_x,k_z) \sim 
\hat{u}^*(k_x,k_z)\iint_{-c Re_\tau/\delta}^{c Re_\tau/\delta}\hat{u}_j(k_x-s_1,k_z-s_2)\frac{\partial \hat{u}(s_1,s_2)}{\partial x_j}, \textrm{d}s_1\textrm{d}s_2,  
\end{equation}
where $c$ is a suitable constant that defines the compact support for $\hat{u}_j$ in the wavenumber space. Furthermore, the integrand itself is expected to be a function of $Re$, as it involves interaction between three pairs of wavenumbers that vary from $O(1/\delta)$ to $O(1/\delta_\nu)$. This suggests that the way $\hat{u}_j$ for $\lambda^+ > 1000$ would feel the wall would be very different from that of laminar flow due to the triadic wave interactions involved. 

The discussion above indicates the complexities on the precise origin of the inactive motions near the wall. Previous studies have shown the existence of an inverse energy transfer from small to large $\lambda$ \cite[]{Kawata.2018,Cho2018,Lee.20196fe}. Given that the wall-normal velocity of these motions is negligible and they are expected to be well correlated below $y^+\simeq 10$ due to their large size, it is tempting to relate their dynamics to two-dimensional turbulence, where inverse energy cascade is known to be an important turbulent energy transport process \cite[e.g.][for a review]{Boffetta2012}. In relation to this, some recent studies have proposed the mechanism of energy transfer from small to large scale \cite[]{Doohan2021,Doohan2022,Ciola2024}. For example, in their numerical simulation where only the two spanwise length scales ($\lambda_z^+=100,200$) were considered, \cite{Doohan2021} showed that the near-wall inverse energy transfer takes place through the subharmonic instability (or transient growth) of the streaks in the buffer layer \cite[]{Kline1967}. An invariant solution of the Navier-Stokes equations underpinning this process was also found \cite[]{Doohan2022}. Taking a step further, \cite{Ciola2024} recently proposed that detuned (or sideband) instabilities (general version of the subharmonic instability) of the buffer-layer streaks could be responsible even for the generation of large-scale structures in the logarithmic and outer regions. However, this scenario is not fully compatible with the simulation of \cite{Doohan2021}, where the time scale of the related inverse energy transfer was found to be much shorter than the typical time scale of energy-containing motions in the log and outer region.

It is also important to note that the inverse transfer of energy from small to large scales is not the only important turbulent transport process in the near-wall region \cite[]{Lee.20196fe}. Indeed, the wall-normal turbulent transport below $y^+\simeq 10$ has been shown to be consistently greater than its inter-scale counterpart (figure \ref{fig:T_filtered}). Furthermore, the wall-normal turbulent transport at $\lambda^+>1000$ is mostly positive for $y^+\lesssim 10$ and monotonically decays as $y\rightarrow 0$ (figure \ref{fig:T_spectra_channel_inner}). This indicates that a great part of the near-wall turbulence at such wavelengths is a consequence of energy transfer towards the wall from above $y^+\approx10$. This observation is also consistent with recent observations on the direct transfer of energy and moment from the outer to the near-wall region \cite[]{Yin2024,Deshpande2024}. Here, it is worth recalling again that the large inactive motions near the wall contain very little wall-normal velocity. This implies that this energy and momentum transfer towards the wall is only possible through nonlinear interactions with the `active' near-wall energy-containing motions at $\lambda^+<1000$ (quasi-streamwise vortices, in particular), as these are the only motions containing the wall-normal velocity near the wall. This feature highlights that the scale interactions depicted by (\ref{eq:4.1}) in turbulent energy transport near the wall are presumably very different from those of two-dimensional turbulence.

\backsection[Acknowledgements]{This research used resources of the Argonne Leadership Computing Facility, a U.S. Department of Energy (DOE) Office of Science user facility at Argonne National Laboratory and is based on research supported by the U.S. DOE Office of Science-Advanced Scientific Computing Research Program, under Contract No. DE-AC02-06CH11357. Y.H. gratefully acknowledges financial support from the European Office of Aerospace Research and Development (FA8655-23-1-7023; Program Manager: Dr D. Smith). }


\backsection[Declaration of interests]{The authors report no conflict of interest.}

\backsection[Data availability statement]{Part of data that support the findings of this study are openly available at https://lee.me.uh.edu. Computational Notebook files are available as supplementary material at \url{https://cocalc.com/share/public_paths/4c1adf25d7ffbbbf79e6b99b048d00b75ca8617d}}

\backsection[Author ORCIDs]{Myoungkyu Lee, 0000-0002-5647-6265; Yongyun Hwang, https://orcid.org/0000-0001-8814-0822}


\appendix
\section{Budget terms in near-wall regions}\label{appA}
\begin{figure}
\centering
  \includegraphics[width=\linewidth]{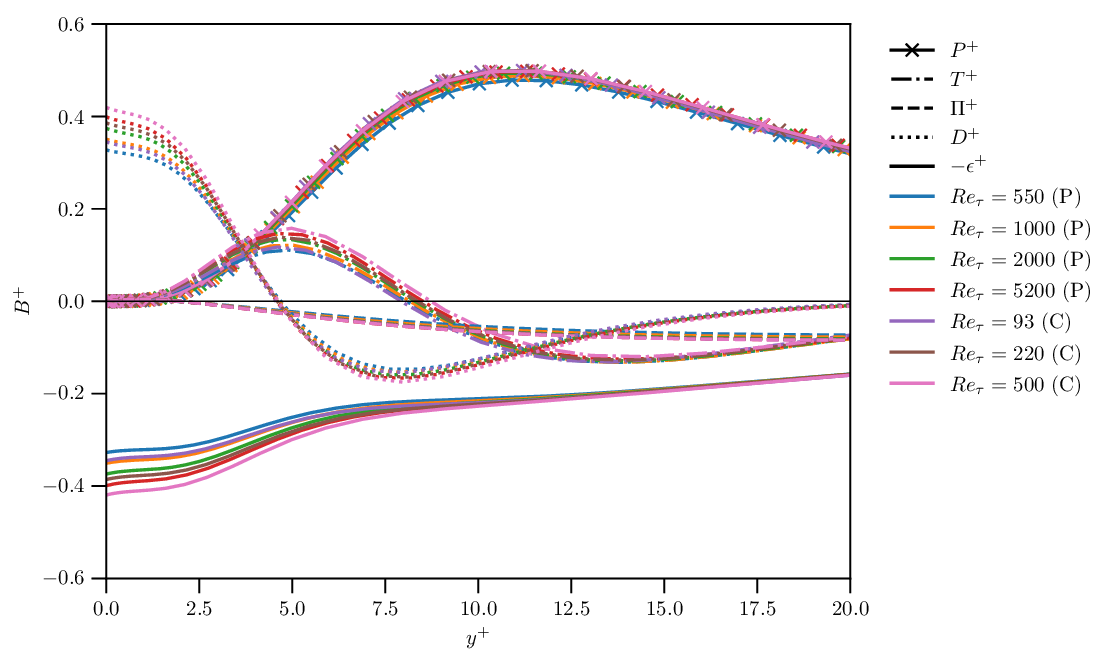}
  \caption{\label{fig:budget_all_unfiltered} budget terms in near-wall regions: P - Poiseuille, C - Couette
}
\end{figure}

The near-wall TKE budget is shown in figure \ref{fig:budget_all_unfiltered} for all $Re_\tau$ considered in both Poiseulle and Couette flow.

\section{Spectra and filtering of streamwise velocity fluctuations}\label{appB}

\begin{figure}
\centering
  \includegraphics[width=\linewidth]{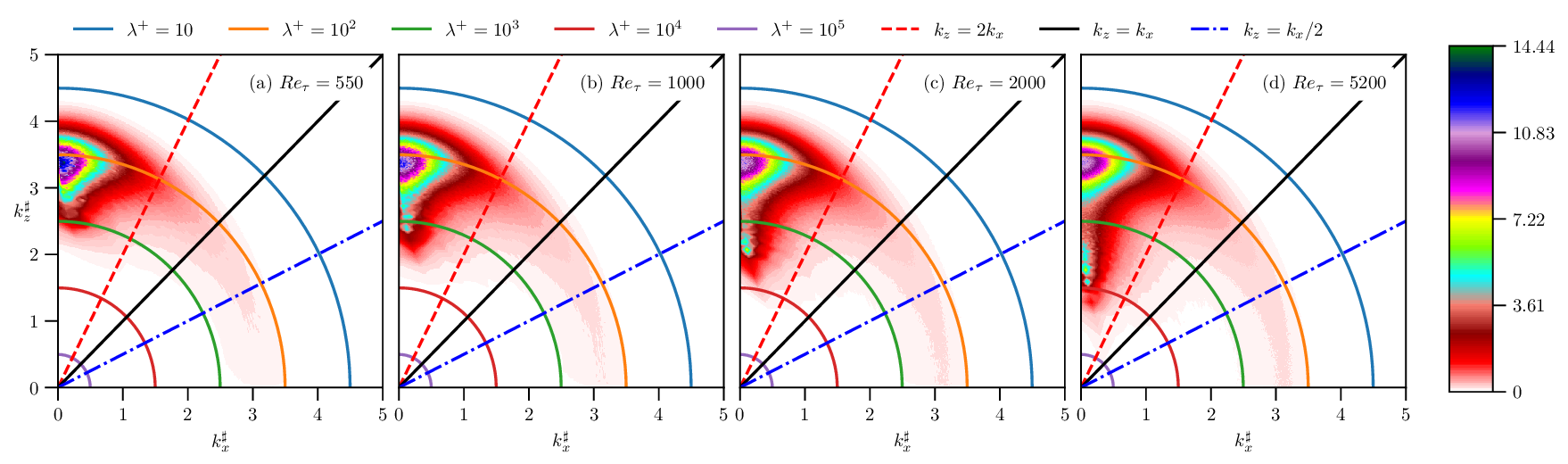}
  \caption{\label{fig:T_spectra_channel_inner_uu_inner} Two-dimensional spectral densities of Poiseuille flows in polar-log coordinates of $u'^2$ at $y^+ = 15$, $k_\mathrm{ref} =1/(50000\delta_v)$.}
\end{figure}

\begin{figure}
\centering
  \includegraphics[width=\linewidth]{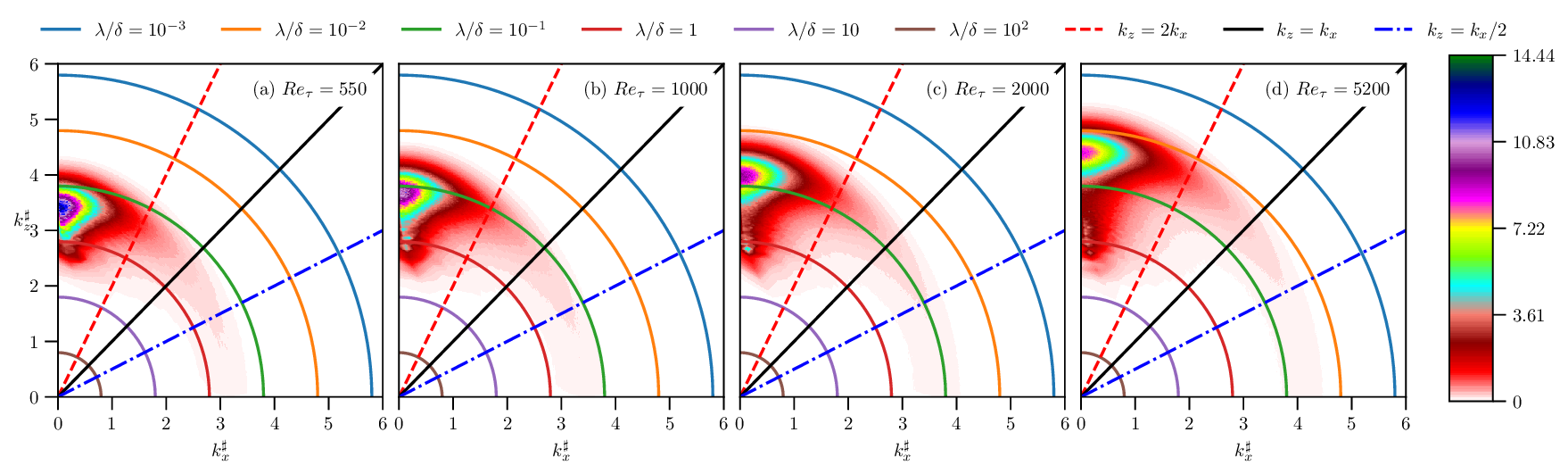}
  \caption{\label{fig:T_spectra_channel_inner_uu_outer} Two-dimensional spectral densities of Poiseuille flows in polar-log coordinates of $u'^2$ at $y^+ = 15$, $k_\mathrm{ref} =1/(100\delta)$.}
\end{figure}

\begin{figure}
\centering
  \includegraphics[width=0.95\linewidth]{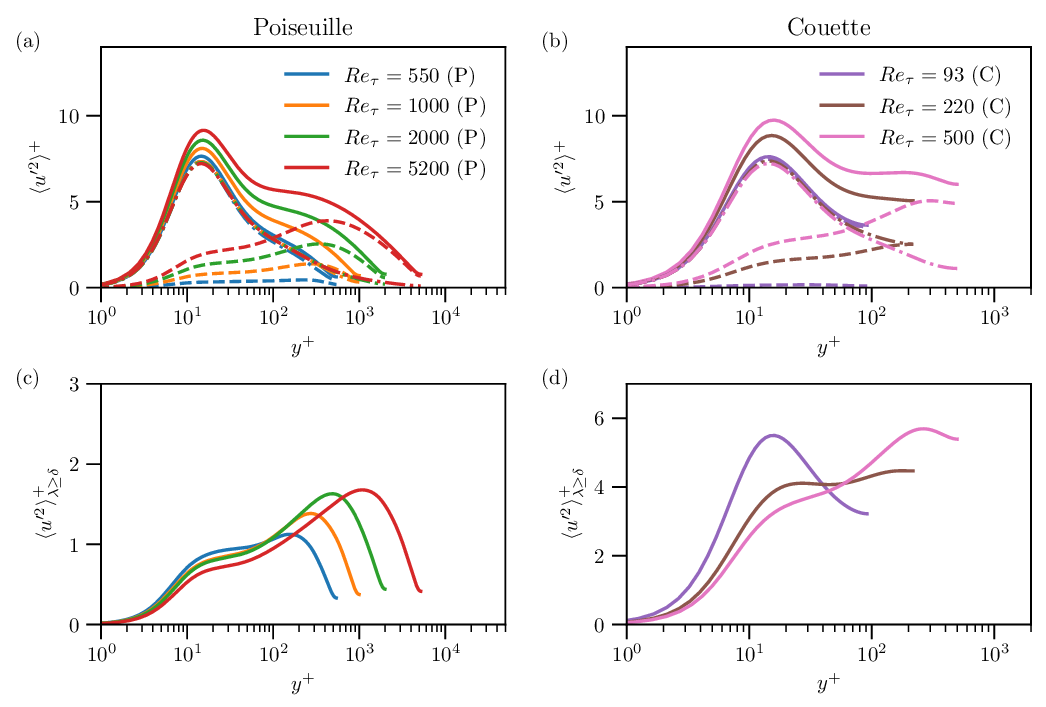}
  \caption{\label{fig:uu} Streamwise velocity variances: 
  (a,b) 
  (------) $\langle u'^2 \rangle^+$,  
  (-- $\cdot$ --) $\langle u'^2 \rangle^+_{\lambda^+<1000}$, 
  (-- -- --), $\langle u'^2 \rangle^+_{\lambda^+\ge1000}$; 
  (c,d) 
  (------) $\langle u'^2 \rangle^+$,  
  (-- $\cdot$ --) $\langle u'^2 \rangle^+_{\lambda<\delta}$, 
  (-- -- --), $\langle u'^2 \rangle^+_{\lambda\ge \delta}$.}
\end{figure}

Figures \ref{fig:T_spectra_channel_inner_uu_inner} and \ref{fig:T_spectra_channel_inner_uu_outer} show the two-dimensional spectral densities of $\langle u'^2 \rangle$, denoted by $E^\#_{u'^2}$, in the polar log coordinate for Poiseulle flow. Overall, the changes in the spectral densities of $E^\#_{u'^2}$ with $Re_\tau$ are very similar to those of $\epsilon_w$ reported in figures \ref{fig:spectra_channel_E_w_inner} and \ref{fig:spectra_channel_E_w_outer}: the spectral density around the peak location at $k_x^\#\simeq 0$ and $\lambda/\delta=1$ (or $k_z^\#\simeq 2.5$) does not exhibit a visible $Re$-dependence, but the region around $k_z\simeq 2k_x$ and $\lambda/\delta=1$ ($k_x^\#\simeq 1.5$ and $k_z^\# \simeq 2.5$) gradually weakens with increasing $Re_\tau$. The spectral densities are also seen to exhibit a good scale separation at a normalised wavelength, $\lambda^+ = 1000$ \cite[see also][]{Lee.20181ab,Lee.20196fe}. 

By applying a high-pass filter with the cut-off threshold $\lambda^+ = 1000$ to the $\langle u'^2 \rangle$ data, a universal behaviour in the small scale is observed like wall shear stress fluctuations in both Poiseuille and Couette flows (figures \ref{fig:uu}a,b), which persists up to $y^+ \approx 70$ for various $Re$s. We note that the identified filtering threshold would be applicable to other canonical wall-bounded turbulent flows, such as pipe \citep{Ahn.2015,Pirozzoli.2021vy8,Yao.2023} and boundary layer flows \citep{Sillero.2013,Samie.2018}, with the expectation of producing the same turbulence statistics on the small scale. Finally, when a low-pass filter with the cut-off threshold $\lambda = \delta$, $\langle u'^2 \rangle$ near the wall ($y^+\lesssim 10$) slowly decays with $Re_\tau$ (figures \ref{fig:uu}c,d), consistent with the behaviour observed in the filtered wall dissipation, $\epsilon_w$ (figures \ref{fig:Euu_wall}c,d).

\nocite{*}
\bibliographystyle{jfm}
\bibliography{reference}

\end{document}